\documentclass[twocolumn,showpacs,prd,floatfix,axodraw]{revtex4}
\usepackage{mathrsfs}
\usepackage{graphicx,,booktabs,bm}
\usepackage{overpic}
\usepackage{color}
\usepackage{amssymb}

\usepackage{mathrsfs,bm,amsmath,amssymb}
\usepackage{longtable,lscape}
\usepackage{txfonts}
\usepackage{amssymb}
\usepackage{indentfirst}
\usepackage{graphicx,,booktabs}
\usepackage{multirow}
\usepackage{color}
\usepackage{amssymb}

\definecolor{cover}{rgb}{0.77,0.87,0.88}
\definecolor{blueone}{rgb}{0.1,0.1,.7}
\definecolor{citec}{rgb}{0.14,0.47,0.09}
\definecolor{two}{rgb}{0.0,0.5,0.}
\definecolor{three}{rgb}{.5,.1,0.15}
\usepackage[bookmarks=true,bookmarksopen=false,plainpages=false,breaklinks=true,
   bookmarksnumbered=true,hypertexnames=false,
   filecolor=blue,urlcolor=three,menucolor=three,
   linkcolor=three,citecolor=blueone, colorlinks,
   anchorcolor=blue,runcolor=pink,frenchlinks=red
   pdfstartview=FitH,pdftitle=title,%
   pdfauthor=author]{hyperref}

\begin{document}
\title{{{Searching for strange hidden-charm pentaquark state $P_{cs}(4459)$ in $\gamma{}p\to{}K^{+}P_{cs}(4459)$ reaction}}}

\author{Cai Cheng$^{1}$}
\author{Feng Yang$^{2}$}
\author{Yin Huang$^{2,3}$}
\email{huangy2019@swjtu.edu.cn;yin.huang@apctp.org}

\affiliation{$^1$ School of Physics and Electronic Engineering, Sichuan Normal University, Chengdu 610101, China}
\affiliation{$^2$ School of Physical Science and Technology, Southwest Jiaotong University, Chengdu 610031,China}
\affiliation{$^3$ Asia Pacific Center for Theoretical Physics,
Pohang University of Science and Technology, Pohang 37673, Gyeongsangbuk-do,
South Korea.}

\date{\today}
\begin{abstract}
We investigate the possibility of studying the strange hidden-charm pentaquark state $P_{cs}(4459)$
by photon-induced reactions on a proton target in an effective Lagrangian approach.  The production process
is described by the $t$-channel $K^{-}$ exchange, the $u$-channel $\Lambda$ exchange, the contract term, and the
$s$- channel nucleon pole.   Our theoretical approach is based on the assumption that $P_{cs}(4459)$ with $J^{P}=1/2^{-}$ or $J^{P}=3/2^{-}$
can be interpreted as a molecule composed of $\bar{D}^{*}\Xi_c$.  Using the coupling constants
of the $P^{J^P}_{cs}$ to $\gamma{}\Lambda$ and $K^{-}p$ channels obtained from molecule picture of the $P^{J^{P}}_{cs}(4459)$,
the total cross-sections of the process $\gamma{}p\to{}P^{J^P}_{cs}K^{+}$ is evaluated.  Our calculation indicates that the
cross-section for $\gamma{}p\to{}P^{1/2^{-}}_{cs}K^{+}$ and $\gamma{}p\to{}P^{3/2^{-}}_{cs}K^{+}$ are of the order of 10.0 pb
and 5.0 pb, respectively.  In addition, we compute the cross-section by assuming $P_{cs}(4459)$ as a compact pentaquark
and find it is quite different from the results of $\bar{D}^{*}\Xi_c$ molecule.
 Those results can be measured in future experiments, such as the Electron-Ion Collider in China
and the United States.  And can be used to test the nature of the $P_{cs}$.
\end{abstract}


\maketitle
\section{INTRODUCTION}
In decades, more and more hadronic exotic states were observed following the accumulation of the precise
data in high energy experiments~\cite{ParticleDataGroup:2020ssz}.  These hadrons have an internal structure
more complex than the simple $\bar{q}q$ configuration for mesons or the $qqq$ configuration for baryons in the traditional
picture of the constituent quark models.  Studying the exotic hadron states is not only conducive to the development
of the hadron spectrum but also provides an important opportunity for us to better understand the strong interaction.

Very recently,  the LHCb experiment reported a new hadronic exotic state, namely $P_{cs}(4459)$, in the $J/\psi{}\Lambda$
invariant mass distributions of the $\Xi_b^{-}\to{}J/\psi{}\Lambda{}K^{-}$ decay~\cite{LHCb:2020jpq}.  The mass and width
of the $P_{cs}(4459)$ are measured to be
\begin{align}
M&=4458.8\pm{}2.9^{+4.7}_{-1.1}~~~{\rm MeV},\nonumber\\
\Gamma&=17.3\pm{}6.5^{+8.0}_{-5.7}~~~~~ {\rm MeV},
\end{align}
respectively.  From the $J/\psi\Lambda$ decay mode, the new structures $P_{cs}(4459)$ contain at least five valence
quarks with isospin is zero.  Because the quark components of $J/\psi$ meson and $\Lambda$ baryon are $\bar{c}c$
and $uds$, respectively,  the $P_{cs}(4459)$ is another new candidate of hidden-charm pentaquark states following
the previous discovery of three hidden-charm pentaquark states~\cite{LHCb:2015yax,Aaij:2019vzc}.
However, its spin-parity quantum number was not confirmed since the statistics is not large enough.

The discovery of the first strange hidden-charm pentaquark immediately intrigues an active discussion on its structure.
Among the theoretical pictures in the field, the molecular picture is a competitive one to explain existing candidates
of exotic states.  The idea comes from the molecular state interpretation of the deuteron, as the deuteron mass is a little
below the corresponding threshold and exhibit a sizable spatial extension.  It immediately leads to a conclusion that a
molecular state is close to the threshold of constituent hadrons.  This feature can be used for defining a hadronic
molecule.  Along this line, one can find that the mass difference between the $P_{cs}(4459)$ and $\bar{D}^{*}\Xi_c$ threshold
is about 19 MeV, which indicates the $P_{cs}(4459)$ could be a candidate for the $\bar{D}^{*}\Xi_c$ molecular state.

Indeed, the QCD sum rules support its interpretation as the $\bar{D}^{*}\Xi_c$ hadronic molecular state of either $J^P=1/2^{-}$
or $3/2^{-}$~~\cite{Chen:2020uif}.  Using the coupled channel unitary approach combined with heavy quark spin and local
hidden gauge symmetries, Ref.~\cite{Xiao:2021rgp} find a pole of $4459.07+i6.89$ MeV below the $\bar{D}^{*}\Xi_c$
threshold consistent with the mass and width of the $P_{cs}(4459)$ state.  In Ref.~\cite{Peng:2020hql}, the
$P_{cs}(4459)$ was regarded as a $\bar{D}^{*}\Xi_c$ molecular pentaquark state with $J^P=3/2^{-}$, or possibly $J^P=1/2^{-}$
with more uncertainties about its mass.   The partial decay width of the molecular $P_{cs}(4459)$ into $J/\psi\Lambda$ is
predicted to be larger for the $J^{P}=3/2^{-}$  configuration than the $J^{P}=1/2^{-}$ case,  in agreement with the
conclusions in Refs.~\cite{Xiao:2021rgp,Yang:2021pio}.  With the quasipotential Bethe-Salpeter equation approach, Ref.~\cite{Zhu:2021lhd} assigned the
$P_{cs}(4459)$ state to the $\bar{D}^{*}\Xi_c$ molecular state with $J^P=3/2^{-}$.  By using a one-boson-exchange model, Ref.~\cite{Chen:2020kco}
concluded that the $P_{cs}(4459)$ state is not a pure $\bar{D}^{*}\Xi_c$ molecular state.   This is the same with our result~\cite{Yang:2021pio}
that the $P_{cs}(4459)$ can be explained as $S$-wave coupled molecular state with  $J^P=3/2^{-}$.  We also proposed that a pure $\bar{D}^{*}\Xi_c$
molecular state with mass about $4459$ and $J^{P}=1/2^{-}$ could exist, and it mainly decays to $D\Xi_c^{'}$ final state.

Its properties, however, such as the spectroscopy and the decay width, can be well explained in the context of the multi-quark
state~\cite{Wang:2015wsa,Azizi:2021utt,Ozdem:2021ugy} with the conclusion that the $P_{cs}(4459)$ can be assigned as hidden charm
compact pentaquark state with $J^P=1/2^{-}$ or $J^P=3/2^{-}$.   We also noted that before the LHCb observation~\cite{LHCb:2020jpq},
the $P_{cs}(4459)$ state mass has been calculated in Ref. \cite{Santopinto:2016pkp} with an extension of the G\"ursey and Radicati 
mass formula \cite{Gursey:1964htz}.   In this paper the authors suggested also to search for $P_{cs}$ states in the
$\Xi_b^{-}\to{} J/\Psi\Lambda K^{-}$ channel and calculated the $P_{cs}(4459)\to{}J/\Psi \Lambda$ strong partial decay
widths\cite{Santopinto:2016pkp} with assumption that $P_{cs}(4459)$ is a compact pentaquark state.

An urgent question of high relevance is to understand the nature of this state: how to distinguish the various interpretations.
One way to distinguish the various interpretations of the $P_{cs}(4459)$ is to study its production processes.  The present knowledge
about the $P_{cs}(4459)$ was obtained from the $pp$ collision~\cite{LHCb:2020jpq}.  High energy photon beams are available at Electron-Ion
Collider in China (EicC)~\cite{Anderle:2021wcy} or in the United States (US-EIC)~\cite{Accardi:2012qut}, which provide another alternative to studying
$P_{cs}(4459)$.  Thus, it will be helpful to understand the nature of the $P_{cs}(4459)$ if we can observe this state in $\gamma{}p\to{}P_{cs}(4459)^{0}K^{+}$
production processes.  The production of the $P_{cs}(4459)$ via a kaon-induced reaction on a nucleon target was discussed in Ref.~\cite{Clymton:2021thh}.

This paper is organized as follows. In Sec.~\ref{Sec: formulism}, we will present the theoretical formalism. In Sec.~\ref{Sec: results}, the
numerical result will be given, followed by discussions and conclusions in the last section.  The Appendix contains technical details  
to compute the partial decay widths of $P_{cs}\to\gamma{}\Lambda$ and $P_{cs}\to{}K^{-}P$ reactions.

\section{THEORETICAL FORMALISM}\label{Sec: formulism}
\subsection{$P_{cs}$ production as $\bar{D}^{*}\Xi_c$ molecule}
\begin{figure}[h!]
\begin{center}
\includegraphics[bb=50 480 750 710, clip, scale=0.50]{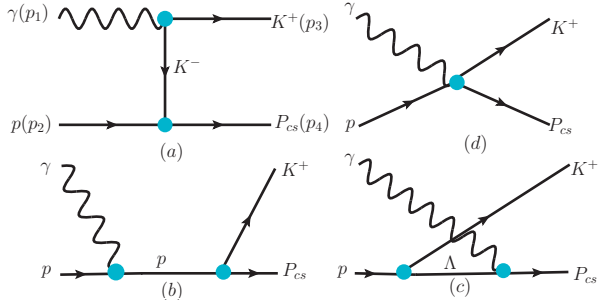}
\caption{Feynman diagrams for the process $\gamma{}p\to{}K^{+}P_{cs}(4459)$. The contributions from the $t$-channel $K^{-}$
exchange (a), $s$-channel nucleon pole (b), $u$-chennel $\Lambda$ hadron (c), and contact term (d) are considered.
In the first diagram, we also show the definition of the kinematics $(p_1, p_2, p_3, p_4)$ that we
use in the present calculation.}
\label{cc1}
\end{center}
\end{figure}

In this work,  we study the process $\gamma{}p\to{}P_{cs}(4459)^{0}K^{+}$ within the effective Lagrangian approach, which has been widely
employed to investigate photoproduction processes.   The relevant Feynman diagrams for the process $\gamma{}p\to{}P_{cs}(4459)^{0}K^{+}$
are depicted in Fig.~\ref{cc1}.   Here we take into account the nucleon-pole contribution in the $s$-channel, the $\Lambda$-pole
contribution in the $u$-channel, and $K$ exchanges in the $t$-channel.   To  ensure  the  gauge  invariance  of  the  total amplitudes,
the contact diagram must be included.

To compute the amplitudes of the diagrams shown in Fig.~\ref{cc1}, we need the effective Lagrangian densities
for the relevant interaction vertices.  The spin parity of the $P_{cs}(4459)$ state was still not determined in
experiments.  The theoretical studies suggested that possible assignments of spin parity of the $P_{cs}(4459)$ are
$J^P=1/2^{-}$ and $3/2^{-}$~\cite{Chen:2020uif,Xiao:2021rgp,Peng:2020hql,Yang:2021pio,Zhu:2021lhd,Chen:2020kco,Wang:2015wsa,Azizi:2021utt,Ozdem:2021ugy}.
In this work, we will consider these two possibilities.  Taking into account these different quantum numbers, we can express the
interactions by the effective Lagrangians~\cite{Clymton:2021thh,Huang:2016tcr,Wang:2015jsa}
\begin{align}
{\cal{L}}_{KNP_{cs}}^{1/2^{-}}&=-g_{KNP_{cs}}\bar{P}_{cs}NK+H.c.,\label{eq1}\\
{\cal{L}}_{KNP_{cs}}^{3/2^{-}}&=-\frac{g_{KNP_{cs}}}{m_Nm_{P_{cs}}}\epsilon^{\mu\nu\alpha\beta}\partial_{\mu}\bar{P}_{cs,\nu}\gamma_{\alpha}N\partial_{\beta}K+H.c.,\label{eq2}\\
{\cal{L}}^{1/2^{-}}_{\gamma\Lambda{}P_{cs}}&=\frac{eh}{2m_{\Lambda}}\bar{\Lambda}\sigma_{\mu\nu}\partial^{\nu}{\cal{A}}^{\mu}P_{cs}+H.c.,\label{eq3}\\
{\cal{L}}^{3/2^{-}}_{\gamma\Lambda{}P_{cs}}&=-e(\frac{ih_1}{2m_{\Lambda}}\bar{\Lambda}\gamma^{\nu}+\frac{h_2}{(2m_{\Lambda})^2}\partial^{\nu}\bar{\Lambda}){\cal{F}}_{\mu\nu}P^{\mu}_{cs}+H.c.,\label{eq4}\\
{\cal{L}}_{\gamma{}KK}&=-ie(K^{-}\partial^{\mu}K^{+}-K^{+}\partial^{\mu}K^{-}){\cal{A}}_{\mu},\label{eq5}\\
{\cal{L}}_{\gamma{}pp}&=-e\bar{p}({\cal{A}}\!\!\!/-\frac{\kappa_{p}}{2m_N}\sigma_{\mu\nu}\partial^{\nu}{\cal{A}}^{\mu})p+H.c.,\label{eq6}\\
{\cal{L}}_{KP\Lambda}&=\frac{g_{KN\Lambda}}{m_N+m_{\Lambda}}\bar{p}\gamma^{\mu}\gamma_5\Lambda\partial_{\mu}K^{+}+H.c.,\label{eq7}
\end{align}
with ${\cal{F}}^{\mu\nu}=\partial_{\mu}{\cal{A}}_{\nu}-\partial_{\nu}{\cal{A}}_{\mu}$, $\sigma^{\mu\nu}=\frac{i}{2}(\gamma^{\mu}\gamma^{\nu}-\gamma^{\nu}\gamma^{\mu})$,
and the electromagnetic fine structure constant $\alpha=e^2/4\pi=1/137$.  The anomalous magnetic momentum reads $\kappa_{p}=1.79$, and $\epsilon^{\mu\nu\alpha\beta}$
is the Levi-Civit\`{a} tensor with
$\epsilon^{0123}=1$.  $P_{cs}$, $\Lambda$, $N$, ${\cal{A}}$, and $K$ are the $P_{cs}$ state, $\Lambda$ baryon, nucleon, photon, and $K$ meson fields,
respectively.   $m_{P_{cs}}$, $m_{\Lambda}$, and $m_{N}$ represent the masses of $P_{cs}(4459)$, $\Lambda$, and the nucleon, respectively.

Here, we discuss relevant coupling constants that we need.  First, the coupling constant $g_{KN\Lambda}$ can be determined by
flavor $SU(3)$ symmetry relations, which give $g_{KN\Lambda}=13.4$~\cite{Oh:2007jd,Wang:2017tpe}.  According to the quark components
of $P_{cs}(4459)$ and $\Lambda$, the decay of the $P_{cs}$ state into $\gamma\Lambda$ should perform via the $c\bar{c}$
annihilation.   For the $P_{cs}(4459)$ with $J^P=3/2^{-}$, there are two different coupling structures for the vertex $P_{cs}\Lambda\gamma$.
Their values can be computed by the radiative decay width of $P_{cs}^{3/2^{-}}\to{}\gamma{}\Lambda$, which is obtained from Eq.~(\ref{eq4})
\begin{align}
\Gamma(P_{cs}^{3/2^{-}}&\to{}\gamma{}\Lambda)=\frac{e^2|\vec{p}|^3_{\gamma}}{12\pi{}}\{\frac{h^2_1}{4m^2_{\Lambda}}(3+\frac{m^2_{\Lambda}}{m^2_{P_{cs}}})+(1+\frac{m_{\Lambda}}{m_{P_{cs}}})\nonumber\\
                                            &\times[\frac{h_1h_2m_{P_{cs}}}{8m^3_{\Lambda}}(3+\frac{m_{\Lambda}}{m_{P_{cs}}})+\frac{h_2^2m^2_{P_{cs}}}{16m^4_{\Lambda}}(1+\frac{m_{\Lambda}}{m_{P_{cs}}})]\},
\end{align}
where $|\vec{p}|_{\gamma}$ is the photon three momenta in the center of mass frame.

As argued in Refs.~\cite{Huang:2016tcr,Wang:2015jsa}, for hidden charm pentaquark state decays into $J/\psi{}p$ the momentum of
the final states are fairly small compared to the nucleon mass.  Thus, the higher partial wave terms proportional to $(p/m_N)^2$
and $(p/m_N)^3$ can be neglected.  It means that the value of the coupling constant $h_2$ related to the higher partial wave term
is zero.   In this work we will only consider the leading order $s$-wave $P^{3/2^{-}}_{cs}\Lambda\gamma$ coupling and leave the
higher partial waves to further studies by following the same pattern in Refs.~\cite{Huang:2016tcr,Wang:2015jsa}.
Thus, we can relate $h_1$ to the radiative decay width of $P_{cs}^{3/2^{-}}\to{}\gamma{}\Lambda$
\begin{align}
\Gamma(P_{cs}^{3/2^{-}}&\to{}\gamma{}\Lambda)=\frac{e^2|\vec{p}|^3_{\gamma}}{12\pi{}}\frac{h^2_1}{4m^2_{\Lambda}}(3+\frac{m^2_{\Lambda}}{m^2_{P_{cs}}}).
\end{align}
However, only one coupling structure exist for the $P_{cs}(4459)$ with $J^P=1/2^{-}$.  With the help of Eq.~(\ref{eq3}),
the coupling constant $h$ can be determined by the radiative decay width of $P_{cs}^{1/2^{-}}\to{}\gamma{}\Lambda$
\begin{align}
\Gamma(P_{cs}^{1/2^{-}}\to{}\gamma{}\Lambda)&=\frac{e^2h^2}{4\pi{}m^2_{\Lambda}}|\vec{p}|^3_{\gamma}.
\end{align}

The coupling constants of $g_{KNP_{cs}}$ are needed as well in our calculation.   The decay processes of $P_{cs}\to{}K^{-}p$
are calculated and the relevant coupling constants $g_{KNP_{cs}}$ can be obtained from their partial decay widths with different
$J^P$ assignments of the $P_{cs}$ states.  The decay rates read
\begin{align}
\Gamma(P_{cs}^{1/2^{-}}\to{}K^{-}p)=\frac{g^2_{KNP_{cs}}}{8\pi}\frac{(m_{P_{cs}}+m_N)^2-m^2_{K^{-}}}{m^2_{P_{cs}}}|\vec{p}|_{K^{-}},\\
\Gamma(P_{cs}^{3/2^{-}}\to{}K^{-}p)=\frac{g^2_{KNP_{cs}}}{24\pi}\frac{(m_{P_{cs}}-m_N)^2-m^2_{K^{-}}}{m_N^2m^2_{P_{cs}}}|\vec{p}|^3_{K^{-}},
\end{align}
where $M_{K^{-}}$ is the masses of the $K^{-}$ meson and $|\vec{p}|_{K^{-}}$ is the three-momenta of the decay products in the
center of mass frame.

In evaluating the production amplitudes of the $\gamma{}p\to{}K^{+}P_{cs}$ reaction, we need to include the form factors
because hadrons are not pointlike particles.  For the $t$-channel $K^{-}$ meson exchange diagram, we take
the form factor as~\cite{Wang:2017tpe}
\begin{align}
{\cal{F}}_{M}(q_{ex},m_{ex})=[\frac{\Lambda^2_{M}-m^2_{ex}}{\Lambda^2_{M}-q_{ex}^2}]^{m}.
\end{align}
For $s$- and $u$-channel diagrams,  we adopt the form factor~\cite{Oh:2007jd,Wang:2017tpe}
\begin{align}
{\cal{F}}_{B}(q_{ex},m_{ex})=[\frac{n\Lambda^4_B}{n\Lambda_B^4+(q_{ex}^2-m_{ex}^2)^2}]^n,
\end{align}
which approaches a Gaussian form as $n\to\infty$.  $q_{ex}$ and $m_{ex}$ are the four-momentum and the mass of the
exchanged particle, respectively.  $\Lambda_{M}$, $\Lambda_B$, $m$ and $n$ will be taken as parameters
and discussed later.

With the vertices Lagrangian densities described in Eqs.~(\ref{eq1})-(\ref{eq7}),  we can further work out the scattering
amplitudes of the $\gamma{}p\to{}K^{+}P_{cs}$ reaction
\begin{align}
{\cal{M}}_{a}^{1/2^{-}}&=-ieg^{1/2^{-}}_{KNP_{cs}}\bar{u}(p_4,s_{cs})u(p_2,s_2)\frac{1}{q_t^2-m^2_{K^{-}}}\nonumber\\
                       &\times(p^{\mu}_{3}-q_t^{\mu})\epsilon_{\mu}(p_1,s_1){\cal{F}}_{K^{-}}(q_t),\\
{\cal{M}}_{b}^{1/2^{-}}&=ieg^{1/2^{-}}_{KNP_{cs}}\bar{u}(p_4,s_{cs})\frac{q\!\!\!/_s+m_p}{q_s^2-m^2_{p}}[\gamma^{\mu}-\frac{\kappa_p}{4m_p}(\gamma^{\mu}p\!\!\!/_1-p\!\!\!/_1\gamma^{\mu})]\nonumber\\
                       &\times{}u(p_2,s_2)\epsilon_{\mu}(p_1,s_1){\cal{F}}_{N}(q_s),\\
{\cal{M}}_{c}^{1/2^{-}}&=-\frac{ehg^{1/2^{-}}_{KN\Lambda}}{4m_{\Lambda}(m_p+m_{\Lambda})}\bar{u}(p_4,s_{cs})(\gamma^{\mu}p\!\!\!/_1-p\!\!\!/_1\gamma^{\mu})\nonumber\\
                       &\times{}\frac{q\!\!\!/_u+m_{\Lambda}}{q^2_{u}-m^2_{\Lambda}}p\!\!\!/_3\gamma_5u(p_2,s_2)\epsilon_{\mu}(p_1,s_1){\cal{F}}_{\Lambda}(q_{\mu}),\\
{\cal{M}}_{d}^{1/2^{-}}&=ieg^{1/2^{-}}_{KNP_{cs}}\bar{u}(p_4,s_{cs}){\cal{C}}^{\mu}_{1/2^{-}}\epsilon_{\mu}(p_1,s_1)u(p_2,s_2)
\end{align}
and
\begin{align}
{\cal{M}}_{a}^{3/2^{-}}&=-i\frac{eg^{3/2^{-}}_{KNP_{cs}}}{m_pm_{P_{cs}}}\epsilon^{\mu\nu\alpha\beta}p_{4\mu}q_{t\beta}(p^{\rho}_{3}-q^{\rho}_{t})\epsilon_{\rho}(p_1,s_1){\cal{F}}_{K^{-}}(q_t)\nonumber\\
                       &\times{}\frac{1}{q^2_{t}-m^2_{K^{-}}}\bar{u}_{\nu}(p_4,s_{cs})\gamma_{\alpha}u(p_2,s_2),\\
{\cal{M}}_{b}^{3/2^{-}}&=-i\frac{eg^{3/2^{-}}_{KNP_{cs}}}{m_pm_{P_{cs}}}\epsilon^{\mu\nu\alpha\beta}p_{4\mu}p_{3\beta}\epsilon_{\rho}(p_1,s_1){\cal{F}}_{N}(q_s)\bar{u}_{\nu}(p_4,s_{cs})\nonumber\\
                       &\times{}\gamma_{\alpha}\frac{q\!\!\!/_s+m_p}{q_s^2-m^2_p}[\gamma^{\rho}-\frac{\kappa_p}{4m_p}(\gamma^{\rho}p\!\!\!/_1-p\!\!\!/_1\gamma^{\rho})]u(p_2,s_2),\\
{\cal{M}}_{c}^{3/2^{-}}&=\frac{eh_1g^{3/2^{-}}_{KNP_{cs}}}{2m_{\Lambda}(m_{\Lambda}+m_p)}\epsilon_{\rho}(p_1,s_1){\cal{F}}_{\Lambda}(q_u)\bar{u}_{\mu}(p_4,s_{cs})\nonumber\\
                       &\times(p^{\mu}_{1}\gamma^{\rho}-p\!\!\!/_1g^{\mu\rho})\frac{q\!\!\!/_u+m_{\Lambda}}{q^2_{u}-m^2_{\Lambda}}p\!\!\!/_3\gamma_5u(p_2,s_2),\\
{\cal{M}}_{d}^{3/2^{-}}&=i\frac{eg^{3/2^{-}}_{KNP_{cs}}}{m_pm_{P_{cs}}}\epsilon^{\mu\nu\alpha\beta}p_{4\mu}\bar{u}_{\nu}(p_4,s_{cs}){\cal{C}}^{\beta\rho}_{3/2^{-}}\gamma_{\alpha}u(p_2,s_2)\nonumber\\
                       &\times\epsilon_{\rho}(p_1,s_1),
\end{align}
where $q_s=p_{1}+p_2=p_3+p_4$, $q_t=p_1-p_3=p_4-p_2$ and $q_{u}=p_2-p_3=p_4-p_1$.  The ${\cal{C}}^{\mu}_{1/2^{-}}$ and
${\cal{C}}^{\beta\rho}_{3/2^{-}}$ are  introduced to ensure that the full photoproduction amplitude satisfies the generalized
Ward-Takahashi identity and thus is fully gauge invariant.  Here, we choose
\begin{align}
{\cal{C}}^{\mu}_{1/2^{-}}&=\frac{2{\cal{F}}_{K^{-}}(q_t)}{q_t^2-m^2_{K^{-}}}p_3^{\mu}-\frac{2{\cal{F}}_{N}(q_s)}{q_s^2-m^2_{p}}p_2^{\mu}\label{eq24},\\
{\cal{C}}^{\beta\rho}_{3/2^{-}}&=\frac{2{\cal{F}}_{K^{-}}(q_t)}{q_t^2-m^2_{K^{-}}}p^{\rho}_{3}q^{\beta}_{t}+\frac{2{\cal{F}}_{N}(q_s)}{q_s^2-m^2_{p}}p_2^{\rho}p^{\beta}_{3}\label{eq25}.
\end{align}

The differential cross section in the center of mass
(c.m.) frame for the process $\gamma{}p\to{}P_{cs}K^{+}$ is calculated using
\begin{align}
\frac{d\sigma}{d\cos\theta}=\frac{m_Nm_{P_{cs}}}{32\pi{}q^2_s}\frac{|\vec{p}^{c.m}_3|}{|\vec{p}^{c.m}_1|}\sum_{s_1,s_2,s_3,s_4}|{\cal{M}}^{J^P=1/2^{-},3/2^{-}}|^2
\end{align}
where ${\cal{M}}^{J^P}={\cal{M}}^{J^P}_{a}+{\cal{M}}^{J^P}_{b}+{\cal{M}}^{J^P}_{c}+{\cal{M}}^{J^P}_{d}$
is the total scattering amplitude of the $\gamma{}p\to{}P_{cs}K^{+}$ reaction.
$\theta$ is the scattering angle of the outgoing $K^{+}$ meson relative to the beam direction, while $\vec{p}^{c.m}_1$
and $\vec{p}^{c.m}_3$ are the photon and $K^{+}$ meson three momenta in the c.m. frame, respectively, which are
\begin{align}
|\vec{p}^{c.m}_1|=\frac{\lambda^{1/2}(q_s^2,0,m_N^2)}{2\sqrt{q^2_s}};~~~|\vec{p}^{c.m}_3|=\frac{\lambda^{1/2}(q_s^2,m^2_{K^{+}},m_{P_{cs}}^2)}{2\sqrt{q^2_s}}
\end{align}
where the $\lambda$ is the K\"{a}llen function with $\lambda(x,y,z)=(x-y-z)^2-4yz$.

\subsection{$P_{cs}$ production as compact pentaquark}
It is helpful if we could estimate the cross-section to make a comparison by assuming $P_{cs}(4459)$ as a compact pentaquark.
Thus, we can judge the different explanations for the structure of $P_{cs}(4459)$ if there exist experimental signals.
Fortunately, a compact pentaquark $P^{0}_{cs}$ with a mass of about $4520\pm{}47$ MeV and isospin $I=0$ is
predicted~\cite{Santopinto:2016pkp}.  And the decay width of this state into $J/\psi\Lambda$  is 7.94 MeV~\cite{Santopinto:2016pkp}.
Considering it as one particle, which is $P_{cs}(4459)$ found in the LHCb experiment\cite{LHCb:2020jpq},  the cross-section
of the process $\gamma{}p\to{}P_{cs}K^{+}$ can be computed with the vector meson dominance mechanism and  relevant Feynman
diagrams plotted in Fig.~\ref{cc2}.
\begin{figure}[h!]
\begin{center}
\includegraphics[bb=50 570 750 705, clip, scale=0.50]{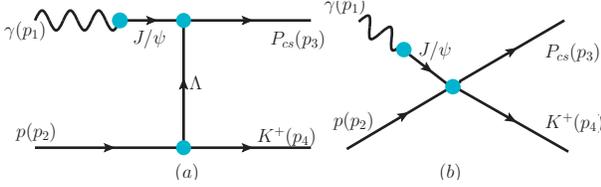}
\caption{Feynman diagrams for the process $\gamma{}p\to{}K^{+}P_{cs}(4459)$ by assuming $P_{cs}(4459)$ as a compact pentaquark.
The contributions from the $t$-channel $\Lambda$ exchange (a) and contact term (b) are considered.
We also show the definition of the kinematics $(p_1, p_2, p_3, p_4)$ that we
use in the present calculation.}
\label{cc2}
\end{center}
\end{figure}
The effective Lagrangian with spin-parity quantum numbers for $P_{cs}$ given~\cite{Clymton:2021thh}
\begin{align}
&{\cal{L}}_{P_{cs}\Lambda{}J/\psi}^{1/2^{-}}=-g_{P_{cs}\Lambda{}J/\psi}\bar{u}_{P_{cs}}\gamma_{\mu}\gamma_5{}u_{\Lambda}J/\psi^{\mu},\\
&{\cal{L}}_{P_{cs}\Lambda{}J/\psi}^{3/2^{-}}=i\frac{g_{P_{cs}\Lambda{}J/\psi}}{2m_{\Lambda}}\bar{u}_{P_{cs},\mu}\gamma_{\nu}(\partial^{\mu}J/\psi^{\nu}-\partial^{\nu}J/\psi^{\mu})u_{\Lambda},
\end{align}
where the $g_{P_{cs}\Lambda{}J/\psi}$ is the coupling constant and can be determined by the decay width of $P_{cs}\to\Lambda{}J/\psi$
\begin{align}
\Gamma(P^{1/2^{-}}_{cs}&\to{}J/\psi\Lambda)=\frac{g^2_{P_{cs}\Lambda{}J/\psi}|\vec{P}|_{J/\psi}}{8\pi{}m^2_{P_{cs}}m^2_{J/\psi}}[m^4_{P_{cs}}+m^2_{P_{cs}}(m^2_{J/\psi}-2m^2_{\Lambda})\nonumber\\
                       &+6m_{P_{cs}}m^2_{J/\psi}m_{\Lambda}-2m^4_{J/\psi}+m^2_{J/\psi}m^2_{\Lambda}+m^4_{\Lambda}],\\
\Gamma(P^{3/2^{-}}_{cs}&\to{}J/\psi\Lambda)= \frac{g^2_{P_{cs}\Lambda{}J/\psi}|\vec{P}|_{J/\psi}}{288\pi{}m^4_{P_{cs}}m^2_{\Lambda}}[3m^6_{P_{cs}}-m^4_{P_{cs}}(m^2_{J/\psi}+5m^2_{\Lambda})\nonumber\\
                       &+12m^3_{P_{cs}}m^2_{J/\psi}m_{\Lambda}-m^2_{P_{cs}}(m^4_{J/\psi}-m^4_{\Lambda})-(m^2_{J/\psi}-m^2_{\Lambda})^3].
\end{align}
where $m_{J/\psi}$ is the mass of $J/\psi$ meson.   Using the corresponding strong decay width $\Gamma{}[P_{cs}\to\Lambda{}J/\psi]=7.94$ MeV
and the masses $m_{P_{cs}}=4458$ MeV,  we obtain $g^{1/2^{-}}_{P_{cs}\Lambda{}J/\psi}=0.299$ and $g^{3/2^{-}}_{P_{cs}\Lambda{}J/\psi}=0.453$.

The effective Lagrangians for the $J/\psi\gamma$ and $\Lambda{}NK$ vertices are expressed as~\cite{Clymton:2021thh,Huang:2013mua}
\begin{align}
{\cal{L}}_{\Lambda{}NK}&=-\frac{f_{\Lambda{}NK}}{m_{\pi}}\bar{\Lambda}\gamma_{\mu}\gamma_5N\partial^{\mu}K+H.c.,\\
{\cal{L}}_{J/\psi\gamma}&=\frac{em^2_{J/\psi}}{f_{J/\psi}}J/\psi_{\mu}{\cal{A}}^{\mu},
\end{align}
where $m_{\pi}=139.57$ MeV is the mass of the $\pi^{+}$ meson and $f_{\Lambda{}NK}=-0.2643$.  There are several ways to determine the
coupling constants $e/f_{J/\psi}$. In this work, we derive the coupling constant $e/f_{J/\psi}=0.0221$ with the experimental
partial decay width $\Gamma_{J/\psi\to{}e^{+}e^{-}}$~\cite{ParticleDataGroup:2020ssz}.

With above details,  the scattering amplitudes of the $\gamma{}p\to{}K^{+}P_{cs}$ reaction can be written as
\begin{align}
{\cal{M}}_{a}^{1/2^{-}}&=i\frac{g_{P_{cs}\Lambda{}J/\psi}f_{\Lambda{}NK}}{m_{\pi}}\frac{e}{f_{J/\psi}}\bar{u}(p_3,s_3)\gamma_{\mu}\gamma_5\frac{q\!\!\!/_t+m_{\Lambda}}{q^2_t-m^2_{\Lambda}}p\!\!\!/_4\gamma_5\nonumber\\
                       &\times{}u(p_2,s_2)(-g^{\mu\nu}+p_1^{\mu}p_1^{\nu}/m^2_{J/\psi})\epsilon_{\nu}(p_1,s_1){\cal{F}}_{\Lambda}(q_t),\\
{\cal{M}}_{a}^{3/2^{-}}&=-i\frac{g_{P_{cs}\Lambda{}J/\psi}f_{\Lambda{}NK}}{2m_{\Lambda}m_{\pi}}\frac{e}{f_{J/\psi}}\bar{u}_{\mu}(p_3,s_3)(p^{\mu}_{1}\gamma^{\eta}-p\!\!\!/_1g^{\mu\eta})\nonumber\\
                       &\times\frac{q\!\!\!/_t+m_{\Lambda}}{q^2_t-m^2_{\Lambda}}p\!\!\!/_4\gamma_5u(p_2,s_2)(-g^{\eta\rho}+p_1^{\eta}p_1^{\rho}/m^2_{J/\psi})\nonumber\\
                       &\times\epsilon_{\rho}(p_1,s_1){\cal{F}}_{\Lambda}(q_t),
\end{align}
The contact term illustrated in Fig.~\ref{cc2}(b) serves to keep the full amplitude gauge invariant.  For the present calculation,
we adopt the form
\begin{align}
{\cal{M}}_{b}^{1/2^{-}}&=-i\frac{g_{P_{cs}\Lambda{}J/\psi}f_{\Lambda{}NK}}{m_{\pi}}\frac{e}{f_{J/\psi}}\frac{2p_1\cdot{}p_3-m_{P_{cs}}+m_{\Lambda}}{q^2_t-m^2_{\Lambda}}\nonumber\\
                       &\times\bar{u}(p_3,s_3)\gamma^{\mu}p\!\!\!/_4u(p_2,s_2)\epsilon_{\mu}(p_1,s_1){\cal{F}}_{\Lambda}(q_t),\\
{\cal{M}}_{b}^{3/2^{-}}&=0.
\end{align}

\section{RESULTS}\label{Sec: results}
According to Refs~\cite{Chen:2020uif,Xiao:2021rgp,Peng:2020hql,Yang:2021pio,Zhu:2021lhd,Chen:2020kco}, $P_{cs}$(4459) may be a molecular state.
However, currently, we cannot fully exclude other possible explanations such as a compact pentaquark state~\cite{Wang:2015wsa, Azizi:2021utt, Ozdem:2021ugy,Santopinto:2016pkp}.
Further research is required to decide whether it is a molecular or compact multi-quark state.   The photon coupling with a quark~\cite{Koniuk:1979vy}
is significantly different from the coupling of the photon to the molecular constituent $\bar{D}^{*}\Xi_c$ of $P_{cs}$(4459)~\cite{Zhu:2020lza,Huang:2021ahp}.
Hence, a precise measurement of the photoproduction is useful to test different interpretations of $P_{cs}$(4459).

We first consider the $P_{cs}$(4459) as pentaquark molecule, its productions in the $\gamma{}p\to{}P_{cs}K^{+}$ reaction is evaluated.
The mechanism including the $t$-channel $K^{-}$ meson exchange, the $u$-channel $\Lambda$ exchange, the contract term, and the
$s$-channel where the nucleon is considered as intermediate state.  To make a reliable prediction for the cross-section of the process
$\gamma{}p\to{}P_{cs}K^{+}$, two issues we need to clarify are, respectively, the relation of the parameters $\Lambda_{M}$, $\Lambda_B$, $m$ and
$n$ to the form factors and the coupling of the $P_{cs}$(4458) with $\gamma{}\Lambda$ and $K^{-}p$.

\begin{figure}[h!]
\begin{center}
\includegraphics[bb=50 300 750 720, clip, scale=0.35]{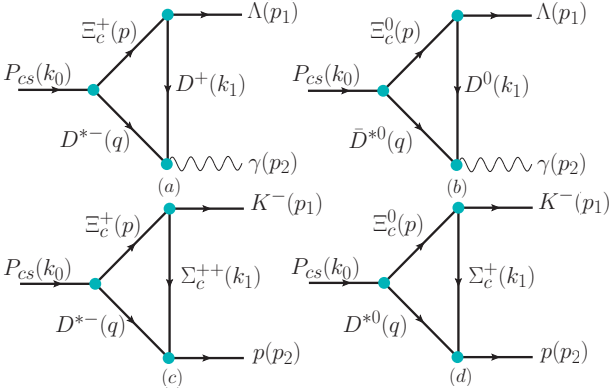}
\caption{Feynman diagrams for $P_{cs}\to{}\gamma{}\Lambda$ and $K^{-}p$ decay processes.
The contributions from the $t$-channel $D$ exchange(a) and $\Sigma_c$ exchange(b) are considered.
We also show the definition of the kinematical $(k_0,p,q,p_1, p_2,k_1)$ that we use in the present
 calculation.}\label{cc2}
\end{center}
\end{figure}
Unfortunately, there is no experimental information on the decay widths for $\Gamma(P_{cs}\to{}\gamma{}\Lambda)$ and
$\Gamma(P_{cs}\to{}K^{-}p)$, as these are very difficult to determine.  Thus, it is necessary to rely on theoretical predictions,
such as those of Ref.~\cite{Yang:2021pio}.  The authors in Ref.~\cite{Yang:2021pio} conclude that the total decay width of the $P_{cs}(4459)$
was reproduced with the assumption that $P_{cs}(4459)$ is a pure $\bar{D}^{*}\Xi_c$ bound state with $J^P=1/2^{-}$,
while in spin-parity $J^P=3/2^{-}$ case may be $S$-wave coupled bound state with lager $\bar{D}^{*}\Xi_c$ component.
Based on the molecular scenario, the partial decay widths of the $P^{}_{cs}(4459)$ with $J^P=1/2^{-}$ and $3/2^{-}$
into $\gamma\Lambda$ and $K^{-}p$ final states through hadronic loops are evaluated with the help of the effective Lagrangians.
The loop diagrams are shown in Fig.~\ref{cc2}.  The obtained partial decay widths are listed in Table~\ref{table1} (more details can be found in the Appendix).
With these decay widths, the coupling constants can be obtained, as in Table~\ref{table1}.
\begin{table}[h!]
\centering
\caption{ The values of the partial decay widths and coupling constants for different $J^{P}$ states.
}\label{table1}
\begin{tabular}{cccccccc}
\hline\hline
    Decay model                 ~~~~~~~&$J^{P}$           ~~~~~~~& $\Gamma$(KeV)           ~~~~~& Coupling constants        ~~     \\\hline
~~$P_{cs}(4459)\to\gamma\Lambda$  ~~~~~~~&$1/2^{-}$         ~~~~~~~& $63.83$                 ~~~~~&$h=0.035$                ~~     \\
~~$P_{cs}(4459)\to\gamma\Lambda$  ~~~~~~~&$3/2^{-}$         ~~~~~~~& $34.18$                 ~~~~~&$h_1=0.05,h_2=0$         ~~    \\
~~$P_{cs}(4459)\to{}K^{-}p$       ~~~~~~~&$1/2^{-}$         ~~~~~~~& $2.05$                  ~~~~~&$g_{KNP_{cs}}=0.0041$    ~~     \\
~~$P_{cs}(4459)\to{}K^{-}p$       ~~~~~~~&$3/2^{-}$         ~~~~~~~& $0.24$                  ~~~~~&$g_{KNP_{cs}}=0.0017$    ~~     \\ \hline
\hline
\end{tabular}
\end{table}

It is worth noting that the decay width of the process $P_{cs}\to\gamma{}\Lambda$ is larger than that of the process $P_{cs}\to{}K^{-}P$.
A possible explanation for this may be that the decay of $P_{cs}$ into $\gamma\Lambda$ should be
easier via $c\bar c$ annihilation than via the Okubo-Zweig-Iizuka mechanism existing in the process $P_{cs}\to{}K^{-}P$.
The larger $P_{cs}\to\gamma{}\Lambda$ decay in Fig.~\ref{cc2} also can be understood due to the fact that the $D$-meson exchange plays
the main role compared to the $\Sigma_c$-baryon exchange.  This mainly originates from the idea that the meson exchange
plays an indispensable role compared to the baryon exchanges in the hadrons interaction.  Therefore, only $\pi$-meson
exchange contribution is allowed in studying the nuclear force~\cite{yua:2001ce}.

At present, the parameters $\Lambda_{M}$, $\Lambda_B$, $m$ and $n$ could not be determined by first principles.  They are usually
determined from the experimental branching ratios.  The free parameters $\Lambda_M$, $\Lambda_B$, $m$ and $n$ are fixed by fitting
the experimental data of the process $\gamma{}p\to{}K^{*+}\Lambda$~\cite{CLAS:2013qgi}, by procedures illustrated in Ref.~\cite{Wang:2017tpe}.
In this work, we adopt the values $\Lambda_{M}=[1.0,1.019,0.993,1.030,1.018]$ GeV, $\Lambda_B=0.9$ GeV and $m=n=2$ because these value
are determined from the experimental data of Ref.~\cite{CLAS:2013qgi} within the same $K^{-}$, $p$ and $\Lambda$ form factors adopted
in the current work.

\begin{figure}[htbp]
\begin{center}
\includegraphics[bb=50 98 750 420, clip,scale=0.50]{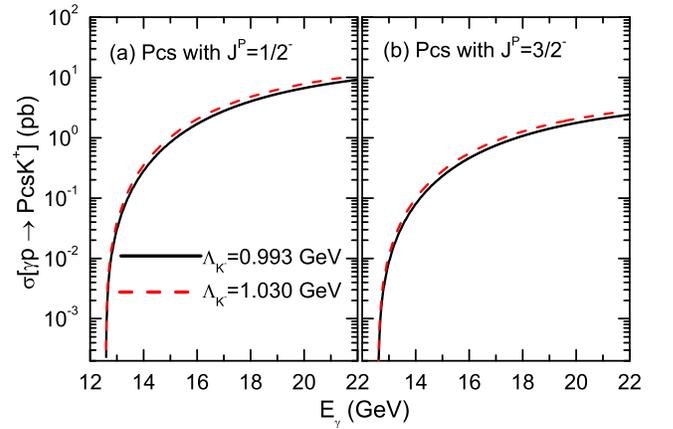}
\caption{(Color online) The total cross section for the process $\gamma{}p\to{}P_{cs}K^{+}$ with different $\Lambda_{K^{-}}$. }
\label{coupling-constants}
\end{center}
\end{figure}
Once the model parameters and coupling constants are determined, the total cross section versus the beam momentum of the photon
for $\gamma{}p\to{}P_{cs}K^{+}$ transition can be evaluated.  In Fig.~\ref{coupling-constants},  the total cross section
of process $\gamma{}p\to{}P_{cs}K^{+}$ with different $\Lambda_{K^{-}}$ is presented,  where we restrict
the value of $\Lambda_{K^-}$ by a reasonable range from 0.993 to 1.030 GeV.  We find that the value of the cross
section increases with the increasing of $\Lambda_{K^-}$.  It is worth mentioning that the value of the
cross section is not very sensitive to the model parameter $\Lambda_{K^{-}}$.  To see how much it depends on the cutoff
parameter $\Lambda_{K^{-}}$, as an example we take the cross section at an energy about $E_{\gamma}=$18.0 GeV.
The obtained cross section ranges from 4.02 pb to 4.77 pb for the process $\gamma{}p\to{}P^{1/2^{-}}_{cs}K^{+}$ and from
1.07 pb to 1.27 pb for the process $\gamma{}p\to{}P^{3/2^{-}}_{cs}K^{+}$.   Hence, we only compute the total cross section
of the process $\gamma{}p\to{}P_{cs}K^{+}$ with $\Lambda_{K^{-}}=1.0$ GeV.

\begin{figure}[htbp]
\begin{center}
\includegraphics[bb=-10 220 500 800, clip,scale=0.40]{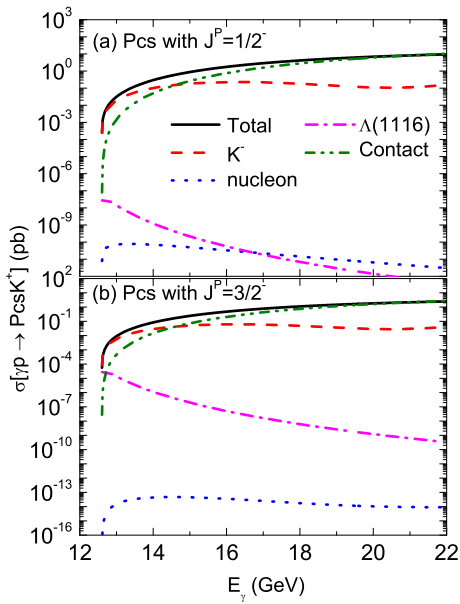}
\caption{(Color online) The cross section for the $\gamma{}p\to{}P_{cs}K^{+}$ reaction
as a function of the beam momentum $E_{\gamma}$ for (a) $P_{cs}$ with $J^P=1/2^{-}$ case
and (b) the $P_{cs}$ with $J^P=3/2^{-}$ case.  The contributions including the $s$-channel nucleon pole (blue dot line),
the $t$-channel $K^{-}$ exchange (red dash line), $\Lambda$(1116) as an intermediate state in the $u$-channel
(magenta dash dot line) and the contact term (olive dash dot dot line).  The black solid line is the total cross section. }
\label{tot-indiv}
\end{center}
\end{figure}
With $\Lambda_{K^{-}}=1.0$ GeV,  the total cross section for the beam momentum $E_{\gamma}$ from reaction threshold up to
22.0 GeV are shown in Fig.~\ref{tot-indiv}.  We find that the total cross section increases sharply near the $K^{+}P_{cs}$
threshold.  At higher energies, the cross section increases continuously but relatively slowly compared to the behavior near
threshold.   With the increase of the beam momentum, the total cross section increases.  The results also show that the
total cross section for $P_{cs}$ production for $J^P=1/2^-$ is larger than for $J^P=3/2^-$.   Taking the
cross section at an energy about 19.00 GeV as example,  the cross section is of the order of 5.515 pb for $P_{cs}^{J^P=1/2^{-}}$
production and 1.464 pb for $P_{cs}^{J^P=3/2^{-}}$ production.  Such a result is very challenging to search for at
EICC~\cite{Anderle:2021wcy} but possible at US-EIC~\cite{Accardi:2012qut} due to a higher luminosity.
We also find that the $t$-channel $K^{-}$ meson exchange plays a predominant role near the threshold, while the contributions
from the contact term become most important when the beam energy $E_{\gamma}$ is larger than 14.63 GeV.  Moreover, the line
shapes of the cross-sections for those two case are the same.

Fig.~\ref{tot-indiv} also tells us that the contributions from
the $s$-channel nucleon pole and $\Lambda$(1116) as an intermediate state in the $u$-channel are small.  The interferences
among them are quite small, especially at high energies, with the consequence that the $t$-channel $K^{-}$ meson exchange and contact term
contributions almost saturate the total cross section.  From Eqs.~\ref{eq24} and \ref{eq25}, we can address that the dominant
contribution from $t$-channel $K^{-}$ exchange and negligible s-channel contribution make the contact term contribution become most
important at high energies.   The dominant $K^{-}$  meson exchange contribution can be easily understood since the $P_{cs}^{1/2^{-}}$
and $P_{cs}^{3/2^{-}}$ are assumed as molecular state with a $\bar{D^{*}}\Xi_c$ component.  Note that the molecule picture~\cite{Chen:2020uif,Xiao:2021rgp,Peng:2020hql,Yang:2021pio,Zhu:2021lhd,Chen:2020kco} is different
from the compact pentaquark picture~\cite{Wang:2015wsa, Azizi:2021utt, Ozdem:2021ugy}.

Our calculation indicates that the contributions from $s$-channel nucleon pole and $u$-channel $\Lambda$(1116) exchange are quite small
and the values are smaller than about the order of $10^{-5}$ pb.   A possible explanation for this may be that the nucleon and
$\Lambda$(1116) are far off the threshold.  It naturally reminds us of what could be the contribution of so many excited states
of the nucleon and $\Lambda$(1116), or other baryonic states in the light quark sector, which can enhance the cross section of
$\gamma{}p\to{}{}P_{cs}K^{+}$ reaction make the EicC easily detect with current luminosity design.
Unfortunately, there is no information on such studies.  Thus, we do not consider the contributions from other states with
heavier mass in this work.

\begin{figure}[htbp]
\begin{center}
\includegraphics[bb=-50 140 500 530, clip,scale=0.50]{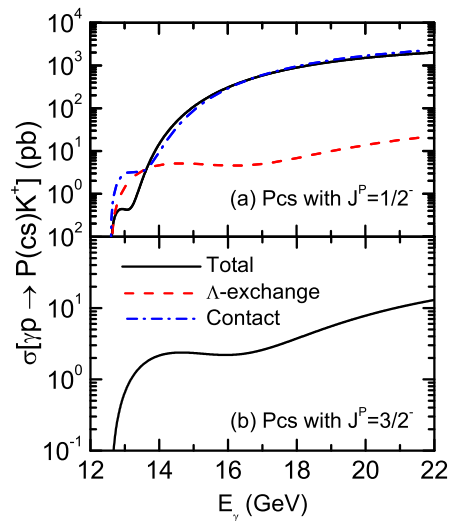}
\caption{(Color online) The cross section for the $\gamma{}p\to{}P_{cs}K^{+}$ reaction
as a function of the beam momentum $E_{\gamma}$ for (a) $P_{cs}$ with $J^P=1/2^{-}$ case
and (b) the $P_{cs}$ with $J^P=3/2^{-}$ case by assuming $P_{cs}(4459)$ as a compact pentaquark.
The contributions including the $t$-channel $\Lambda$ exchange (red dash line) and the contact
term (blue dash dot).  The black solid line is the total cross section. }
\label{tot-indiv-two}
\end{center}
\end{figure}

Now we turn to the cross section for $\gamma{}p\to{}P_{cs}K^{+}$ by assuming $P_{cs}(4459)$ as a compact pentaquark.
The cross section for the beam momentum $E_{\gamma}$ from reaction threshold up to 22.0 GeV are shown in Fig.~\ref{tot-indiv-two}.
We find that the total cross section increases sharply near the  threshold.   At higher energies, the cross section
increases continuously but relatively slowly compared to the behavior near threshold.   With the increase of the beam momentum,
the total cross section increases.  The results also show that the total cross section for $P_{cs}$ production for
$J^P=1/2^-$ is much larger than for $J^P=3/2^-$.  Such as the total cross sections at an energy about $E_{\gamma}=20.0$ GeV can
reach 1509.36 pb for $P_{cs}^{J^P=1/2^{-}}$ production and 7.84 pb for $P_{cs}^{J^P=3/2^{-}}$ production.   Moreover, the contributions
from the contact term for $J^P=1/2^{-}$ plays a predominant role, which almost equal to the total cross section.

Comparing the cross-sections shown in Figs.~\ref{tot-indiv} and ~\ref{tot-indiv-two},
we can find that if the $P_{cs}$ is produced as a compact pentaquark state, the cross-section and the line shapes of the cross-sections are different
from the results that are obtained by assuming $P_{cs}(4459)$ as $\bar{D}^{*}\Xi_c$ bound state.   These differences will
be very useful to help us to test various interpretations of the  $P_{cs}(4459)$.

\section{SUMMARY}
In this paper, we made a detailed exploration of the nonresonant contribution to the $\gamma{}p\to{}P_{cs}K^{+}$ reaction,
intending to find a reasonable estimate of the $P_{cs}$ production rates at relatively high energies, where no data are
available up to now.  The production process is described by the $t$-channel $K^{-}$ exchange, $s$-channel nucleon pole,
$u$-channel $\Lambda$(1116) exchange, and the contact term.   Based on the theoretical conclusion in Ref~\cite{Yang:2021pio}
that $P_{cs}(4459)$ can be interpreted as $\bar{D}^{*}\Xi_c$ bound state by studying the strong decay width,  the coupling
constants of the $P_{cs}$ to $\gamma{}p$ and $K^{-}p$ needed in this work can be studied.  The relevant Feynman diagrams are
shown in Fig.~\ref{cc2} and the results are given in Table~\ref{table1}.  For comparison, the cross section for $\gamma{}p\to{}P_{cs}K^{+}$
is given also by assuming $P_{cs}(4459)$ as a compact pentaquark.

With assuming $P_{cs}$ as a $\bar{D}^{*}\Xi_c$ bound state, the cross section for $\gamma{}p\to{}P^{1/2^{-}}_{cs}K^{+}$ and
$\gamma{}p\to{}P^{3/2^{-}}_{cs}K^{+}$ reactions can reach 10.0 pb and 5.0 pb, respectively.  However, the cross section for $\gamma{}p\to{}P^{1/2^{-}}_{cs}K^{+}$ and
$\gamma{}p\to{}P^{3/2^{-}}_{cs}K^{+}$ reactions can reach 2000.0 pb and 13.0 pb, respectively, by considering $P_{cs}$ as a compact pentaquark.
Although the photoproduction cross-section is quite small, it is enough to
test various interpretations of the  $P_{cs}(4459)$ thanks to the low background of the exclusive and specific reaction proposed in this work.  The future electron-ion colliders (EIC) of high luminosity in China ($2-4\times10^{33}$ cm$^{-2}$s$^{-1}$)~\cite{Anderle:2021wcy} and
the United States (US-EIC) ($10^{34}$ cm$^{-2}$s$^{-1}$)~\cite{Accardi:2012qut} provide a good platform for this purpose.

\section*{Acknowledgments}
Yin Huang want to thanks for
the support by the Development and Exchange Platform for
the Theoretic Physics of Southwest Jiaotong University under Grants No.11947404 and No.12047576,
the Fundamental Research Funds for the Central Universities (Grant No.
2682020CX70), and the National Natural Science Foundation
of China under Grant No.12005177.

\section{Appendix}
In this appendix, we show how to compute the partial decay widths of $P_{cs}\to\gamma{}\Lambda$ and $P_{cs}\to{}K^{-}P$ reactions.
The corresponding Feynman diagrams are shown in Fig.~\ref{cc2}.  To compute the diagrams, we require the effective Lagrangian
densities for the relevant interaction vertices.  The corresponding Lagrangian densities, theoretical formalism and coupling
constants can be found in Ref.~\cite{Yang:2021pio}.  Here, we do not go into details.  The amplitudes for $P_{cs}$ with
$J^P=\frac{1}{2}^-$ case can be shown as
\begin{align}
{\cal{M}}_{a}&=\bar{\mu}(p_1)\bigg(-\frac{e}{4\sqrt{2}}g_{\Xi_{c}^+\Lambda D^+}g_{D^{*-}D^{+}\gamma}g_{p_{c s}}^{1/2}\int\frac{d^{4}k_{1}}{(2\pi)^{4}}\nonumber\\
             &\times\Phi((p\omega_{\bar{D}^{*-}}-q\omega_{\Xi_c^+})^2)\epsilon_{\mu\nu\alpha\beta}\gamma^5\frac{p\!\!\!/+m_{\Xi_{c}^{+}}}{p^2-m_{\Xi_{c}^{+}}^2}\gamma^{\sigma}\gamma^5\nonumber\\
             &\times\frac{-g^{\lambda\sigma}+q^{\lambda}q^{\sigma}/m_{D^{*-}}^2}{q^2-m_{D^{*-}}^2}(p_{2}^{\mu}g^{\theta\nu}-p_2^{\nu}g^{\theta\mu})\nonumber\\
             &\times(q^{\beta}g^{\alpha\lambda}-q^{\alpha}g^{\beta\lambda})\frac{1}{k_{1}^{2}-m_{D^{+}}^2}\bigg)\varepsilon^{\lambda}(p_2)\mu(k_0), \\
{\cal{M}}_{b}&=\bar{\mu}(p_1)\bigg(\frac{e}{4\sqrt{2}}g_{\Xi_{c}^0\Lambda D^0}g_{D^{*0}D^{0}\gamma}g_{p_{c s}}^{1 / 2}\int\frac{d^{4}k_{1}}{(2\pi)^{4}}\nonumber\\
               &\times\Phi((p\omega_{\bar{D}^{*0}}-q\omega_{\Xi_c^0})^2)\epsilon_{\mu\nu\alpha\beta}\gamma^5\frac{p\!\!\!/+m_{\Xi_{c}^{0}}}{p^2-m_{\Xi_{c}^{0}}^2}\gamma^{\sigma}\gamma^5\nonumber\\
               &\times\frac{-g^{\lambda\sigma}+q^{\lambda}q^{\sigma}/m_{D^{*0}}^2}{q^2-m_{D^{*0}}^2}(p_{2}^{\mu}g^{\theta\nu}-p_2^{\nu}g^{\theta\mu})\nonumber\\
               &\times(q^{\beta}g^{\alpha\lambda}-q^{\alpha}g^{\beta\lambda})\frac{1}{k_{1}^{2}-m_{D^{0}}^2}\bigg)\varepsilon^{\lambda}(p_2)\mu(k_0),\\
{\cal{M}}_{c}&=\bar{\mu}(p_2)\bigg(\frac{1}{\sqrt{2}}g_{\Xi_{c}^+\Sigma_{c}^{++} K^-}g_{D^{*-}p\Sigma_{c}^{++}}g_{p_{c s}}^{1 / 2}\int\frac{d^{4}k_{1}}{(2\pi)^{4}}\nonumber\\
               &\times\Phi((p\omega_{\bar{D}^{*-}}-q\omega_{\Xi_c^+})^2)\gamma^{\rho}\frac{k\!\!\!/_{1}+m_{\Sigma_{c}^{++}}}{k_{1}^{2}-m_{\Sigma_{c}^{++}}^2}\nonumber\\
	           &\times
               \gamma^{5}\frac{p\!\!\!/+m_{\Xi_{c}^{+}}}{p^2-m_{\Xi_{c}^{+}}^2}\gamma^{\sigma}\gamma^5\frac{-g^{\lambda\sigma}+q^{\lambda}q^{\sigma}/m_{D^{*-}}^2}{q^2-m_{D^{*-}}^2}
               \bigg)\mu(k_0), \\
{\cal{M}}_{d}&=\bar{\mu}(p_2)\bigg(-\frac{1}{\sqrt{2}}g_{\Xi_{c}^0\Sigma_{c}^{+} K^-}g_{D^{*0}p\Sigma_{c}^{+}}g_{p_{c s}}^{1 / 2}\int\frac{d^{4}k_{1}}{(2\pi)^{4}}\nonumber\\
               &\times\Phi((p\omega_{\bar{D}^{*0}}-q\omega_{\Xi_c^0})^2)\gamma^{\rho}\frac{k\!\!\!/_{1}+m_{\Sigma_{c}^{+}}}{k_{1}^{2}-m_{\Sigma_{c}^{+}}^2}\nonumber\\
               &\times \gamma^{5}\frac{p\!\!\!/+m_{\Xi_{c}^{0}}}{p^2-m_{\Xi_{c}^{0}}^2}\gamma^{\sigma}\gamma^5\frac{-g^{\lambda\sigma}+q^{\lambda}q^{\sigma}/m_{D^{*0}}^2}{q^2-m_{D^{*0}}^2}
               \bigg)\mu(k_0)        
\end{align}
and for $J^P=\frac{3}{2}^-$ have
\begin{align}
{\cal{M}}_{a}&=\bar{\mu}(p_1)\bigg(i\frac{e}{4\sqrt{2}}g_{\Xi_{c}^+\Lambda D^+}g_{D^{*-}D^{+}\gamma}g_{p_{c s}}^{3 / 2}\int\frac{d^{4}k_{1}}{(2\pi)^{4}}\nonumber\\
             &\times\Phi((p\omega_{\bar{D}^{*-}}-q\omega_{\Xi_c^+})^2)\epsilon_{\mu\nu\alpha\beta}\gamma^5\frac{p\!\!\!/+m_{\Xi_{c}^{+}}}{p^2-m_{\Xi_{c}^{+}}^2}\nonumber\\
             &\times \frac{-g^{\lambda\sigma}+q^{\lambda}q^{\sigma}/m_{D^{*-}}^2}{q^2-m_{D^{*-}}^2}(p_{2}^{\mu}g^{\theta\nu}-p_2^{\nu}g^{\theta\mu})\nonumber\\
             &\times(q^{\beta}g^{\alpha\lambda}-q^{\alpha}g^{\beta\lambda})\frac{1}{k_{1}^{2}-m_{D^{+}}^2}\bigg)\varepsilon^{\lambda}(p_2)\mu^{\sigma}(k_0),\\
{\cal{M}}_{b}&=\bar{\mu}(p_1)\bigg(-i\frac{e}{4\sqrt{2}}g_{\Xi_{c}^0\Lambda D^0}g_{D^{*0}D^{0}\gamma}g_{p_{c s}}^{1 / 2}\int\frac{d^{4}k_{1}}{(2\pi)^{4}}\nonumber\\
             &\times\Phi((p\omega_{\bar{D}^{*0}}-q\omega_{\Xi_c^0})^2)\epsilon_{\mu\nu\alpha\beta}\gamma^5\frac{p\!\!\!/+m_{\Xi_{c}^{0}}}{p^2-m_{\Xi_{c}^{0}}^2}\nonumber
\end{align}
\begin{align}
             &\times \frac{-g^{\lambda\sigma}+q^{\lambda}q^{\sigma}/m_{D^{*0}}^2}{q^2-m_{D^{*0}}^2}(p_{2}^{\mu}g^{\theta\nu}-p_2^{\nu}g^{\theta\mu})\nonumber\\
             &\times(q^{\beta}g^{\alpha\lambda}-q^{\alpha}g^{\beta\lambda})\frac{1}{k_{1}^{2}-m_{D^{0}}^2}\bigg)\varepsilon^{\lambda}(p_2)\mu^{\sigma}(k_0), \\
{\cal{M}}_{c}&=\bar{\mu}(p_2)\bigg(-i\frac{1}{\sqrt{2}}g_{\Xi_{c}^+\Sigma_{c}^{++} K^-}g_{D^{*-}p\Sigma_{c}^{++}}g_{p_{c s}}^{1 / 2}\int\frac{d^{4}k_{1}}{(2\pi)^{4}}\nonumber\\
             &\times\Phi((p\omega_{\bar{D}^{*-}}-q\omega_{\Xi_c^+})^2)\gamma^{\rho}\frac{k\!\!\!/_{1}+m_{\Sigma_{c}^{++}}}{k_{1}^{2}-m_{\Sigma_{c}^{++}}^2}\nonumber\\
            &\times \gamma^{5}\frac{p\!\!\!/+m_{\Xi_{c}^{+}}}{p^2-m_{\Xi_{c}^{+}}^2}\frac{-g^{\lambda\sigma}+q^{\lambda}q^{\sigma}/m_{D^{*-}}^2}{q^2-m_{D^{*-}}^2}\bigg)\mu^{\sigma}(k_0),\\
{\cal{M}}_{d}&=\bar{\mu}(p_2)\bigg(i\frac{1}{\sqrt{2}}g_{\Xi_{c}^0\Sigma_{c}^{+} K^-}g_{D^{*0}p\Sigma_{c}^{+}}g_{p_{c s}}^{1 / 2}\int\frac{d^{4}k_{1}}{(2\pi)^{4}}\nonumber\\
              &\times\Phi((p\omega_{\bar{D}^{*0}}-q\omega_{\Xi_c^0})^2)\gamma^{\rho}\frac{k\!\!\!/_{1}+m_{\Sigma_{c}^{+}}}{k_{1}^{2}-m_{\Sigma_{c}^{+}}^2}\nonumber\\
             &\times \gamma^{5}\frac{p\!\!\!/+m_{\Xi_{c}^{0}}}{p^2-m_{\Xi_{c}^{0}}^2}\frac{-g^{\lambda\sigma}+q^{\lambda}q^{\sigma}/m_{D^{*0}}^2}{q^2-m_{D^{*0}}^2}
             \bigg)\mu^{\sigma}(k_0),         
\end{align}
where $\omega_{i}=m_{i}/(m_{i}+m_{j})$.  In the above Lagrangians, the effective correlation function $\Phi(y^2)$ show the distribution of the components
in the hadronic molecule $P_{cs}(4459)$ state.  Moreover, the role of the correlation function $\Phi(y^2)$ also is to avoid the Feynman diagrams
ultraviolet divergence, as the Fourier transform should vanish quickly in the ultraviolet region in the Euclidean space.  We adopt the form as
used in Refs.~\cite{Dong:2008gb,Dong:2017rmg},
\begin{align}
\Phi(-p^2)\doteq{}\exp(-p_E^2/\alpha^2),
\end{align}
where $p_E$ is the Euclidean Jacobi momentum. At present, the experimental total widths of  the $P_{cs}(4459)$
that can be considered as a molecule can be well explained with $\alpha$ = 1.0 GeV~\cite{Yang:2021pio}.
Therefore we take $\alpha$ = 1.0 GeV in this work to compute the partial decay widths of
$P_{cs}\to\gamma{}\Lambda$ and $P_{cs}\to{}K^{-}P$ reactions.

Once the amplitudes are determined, the corresponding partial decay widths can be obtained, which read
\begin{align}
\Gamma(P_{cs}\to)=\int\frac{1}{2J+1}\frac{1}{32\pi^2}\frac{|\vec{p}_1|}{m^2_{P_{cs}}}|\bar{{\cal{M}}}|^2d\Omega,
\end{align}
where the $J$ is the total angular momentum of $P_{cs}(4459)$, $|\vec{p}_1|$ is the three-momenta of the decay
products in the center of mass frame, the overline indicates the sum over the polarization vectors of the final hadrons.
The $\Omega$ is the space angle of the final particle in the rest frame of $P_{cs}(4459)$.


\begin{thebibliography}{90}
\bibitem{ParticleDataGroup:2020ssz}
P.~A.~Zyla \textit{et al.} [Particle Data Group],
PTEP \textbf{2020}, 083C01 (2020).



\bibitem{LHCb:2020jpq}
R.~Aaij \textit{et al.} [LHCb],
Sci. Bull. \textbf{66}, 1278-1287 (2021).



\bibitem{LHCb:2015yax}
R.~Aaij \textit{et al.} [LHCb],
Phys. Rev. Lett. \textbf{115}, 072001 (2015).


\bibitem{Aaij:2019vzc}
R.~Aaij \textit{et al.} [LHCb],
Phys. Rev. Lett. \textbf{122}, 222001 (2019).


\bibitem{Chen:2020uif}
H.~X.~Chen, W.~Chen, X.~Liu and X.~H.~Liu,
Eur. Phys. J. C \textbf{81}, 409 (2021).


\bibitem{Xiao:2021rgp}
C.~W.~Xiao, J.~J.~Wu and B.~S.~Zou,
Phys. Rev. D \textbf{103}, 054016 (2021).


\bibitem{Peng:2020hql}
F.~Z.~Peng, M.~J.~Yan, M.~S\'anchez S\'anchez and M.~P.~Valderrama,
[arXiv:2011.01915 [hep-ph]].


\bibitem{Yang:2021pio}
F.~Yang, Y.~Huang and H.~Q.~Zhu,
[arXiv:2107.13267 [hep-ph]].

\bibitem{Zhu:2021lhd}
J.~T.~Zhu, L.~Q.~Song and J.~He,
Phys. Rev. D \textbf{103}, 074007 (2021).

\bibitem{Chen:2020kco}
R.~Chen,
Phys. Rev. D \textbf{103}, 054007 (2021).

\bibitem{Wang:2015wsa}
Z.~G.~Wang,
Eur. Phys. J. C \textbf{76}, 142 (2016).


\bibitem{Azizi:2021utt}
K.~Azizi, Y.~Sarac and H.~Sundu,
Phys. Rev. D \textbf{103}, 094033 (2021).

\bibitem{Ozdem:2021ugy}
U.~\"Ozdem,
Eur. Phys. J. C \textbf{81}, 277 (2021).


\bibitem{Santopinto:2016pkp}
E.~Santopinto and A.~Giachino,
Phys. Rev. D \textbf{96}, 014014 (2017).


\bibitem{Gursey:1964htz}
F.~Gursey and L.~A.~Radicati,
Phys. Rev. Lett. \textbf{13}, 173-175 (1964).


\bibitem{Anderle:2021wcy}
D.~P.~Anderle, V.~Bertone, X.~Cao, L.~Chang, N.~Chang, G.~Chen, X.~Chen, Z.~Chen, Z.~Cui and L.~Dai, \textit{et al.}
Front. Phys. (Beijing) \textbf{16}, 64701 (2021).

\bibitem{Accardi:2012qut}
A.~Accardi, J.~L.~Albacete, M.~Anselmino, N.~Armesto, E.~C.~Aschenauer, A.~Bacchetta, D.~Boer, W.~K.~Brooks, T.~Burton and N.~B.~Chang, \textit{et al.}
Eur. Phys. J. A \textbf{52}, 268 (2016).

\bibitem{Clymton:2021thh}
S.~Clymton, H.~J.~Kim and H.~C.~Kim,
Phys. Rev. D \textbf{104}, 014023 (2021).


\bibitem{Huang:2016tcr}
Y.~Huang, J.~J.~Xie, J.~He, X.~Chen and H.~F.~Zhang,
Chin. Phys. C \textbf{40}, 124104 (2016).

\bibitem{Wang:2015jsa}
Q.~Wang, X.~H.~Liu and Q.~Zhao,
Phys. Rev. D \textbf{92}, 034022 (2015).

\bibitem{Oh:2007jd}
Y.~Oh, C.~M.~Ko and K.~Nakayama,
Phys. Rev. C \textbf{77}, 045204 (2008).

\bibitem{Wang:2017tpe}
A.~C.~Wang, W.~L.~Wang, F.~Huang, H.~Haberzettl and K.~Nakayama,
Phys. Rev. C \textbf{96}, 035206 (2017).


\bibitem{Koniuk:1979vy}
R.~Koniuk and N.~Isgur,
Phys. Rev. D \textbf{21}, 1868 (1980)
[erratum: Phys. Rev. D \textbf{23}, 818 (1981)].

\bibitem{Zhu:2020lza}
H.~Zhu and Y.~Huang,
Chin. Phys. C \textbf{44}, 083101 (2020).


\bibitem{Huang:2021ahp}
Y.~Huang, F.~Yang and H.~Zhu,
Chin. Phys. C \textbf{45}, 073112 (2021).

\bibitem{yua:2001ce}
H. Yukawa, Proc. Phys. Math. Soc. Jap. 17, 48 (1935), [Prog. Theor. Phys. Suppl.1,1(1935)].


\bibitem{CLAS:2013qgi}
W.~Tang \textit{et al.} [CLAS],
Phys. Rev. C \textbf{87},065204 (2013).


\bibitem{Dong:2008gb}
  Y.~Dong, A.~Faessler, T.~Gutsche and V.~E.~Lyubovitskij,
  Phys.\ Rev.\ D {\bf 77}, 094013 (2008).

\bibitem{Dong:2017rmg}
  Y.~Dong, A.~Faessler, T.~Gutsche, Q.~Lu  and V.~E.~Lyubovitskij,
  Phys.\ Rev.\ D {\bf 96},074027 (2017).

\bibitem{Huang:2013mua}
Y.~Huang, J.~He, H.~F.~Zhang and X.~R.~Chen,
J. Phys. G \textbf{41}, 115004 (2014).




\end{thebibliography}
\end{document}